# Risk assessment of airborne transmission of COVID-19 by asymptomatic individuals under different practical settings

**Authors:** Siyao Shao[1, 2], Dezhi Zhou[1], Ruichen He[1,2], Jiaqi Li[1,2], Shufan Zou[1], Kevin Mallery[1,2], Santosh Kumar[1,2], Suo Yang[1], Jiarong Hong[1,2, *]

**Affiliations:**

[1]Department of Mechanical Engineering, 111 Church ST SE, University of Minnesota, Minneapolis, MN, USA 55414.

[2]Saint Anthony Falls Laboratory, 2 3rd AVE SE, University of Minnesota, Minneapolis, MN, USA 55414.

*Email addresses of the first corresponding author: jhong@umn.edu

**Abstract:** The lack of quantitative risk assessment of airborne transmission of COVID-19 under practical settings leads to large uncertainties and inconsistencies in our preventive measures. Combining *in situ* measurements and computational fluid dynamics simulations, we quantify the exhaled particles from normal respiratory behaviors and their transport under elevator, small classroom, and supermarket settings to evaluate the risk of inhaling potentially virus-containing particles. Our results show that the design of ventilation is critical for reducing the risk of particle encounters. Inappropriate design can significantly limit the efficiency of particle removal, create local hot spots with orders of magnitude higher risks, and enhance particle deposition causing surface contamination. Additionally, our measurements reveal the presence of a substantial fraction of crystalline particles from normal breathing and its strong correlation with breathing depth.

## 1. Introduction

The global pandemic of COVID-19 (caused by the SARS-CoV-2 virus) has demonstrated the extraordinary transmissibility of the virus, with more than 15 million people infected as of writing. However, mechanisms to contain the disease are regionally variable, with vastly different approaches being utilized by different countries, regions (such as US states), and even cities (Anderson et al. 2020). This inconsistency is due in part to a lack of understanding of the transmission pathways of the disease (Anderson et al. 2020). Although it has been well-accepted that the disease can be transmitted through large droplets (>5 µm) capable of carrying sufficient viral load produced by coughing and sneezing (Gralton et al. 2011, World Health Organization 2004), there is substantial debate regarding whether the transmission can be airborne with small droplets (Asadi et al. 2020, Lewis 2020, and Prather et al. 2020). Nevertheless, growing evidence, including the detection of SARS-CoV-2 RNA in collected particles (Liu et al. 2020) and the ability of SARS-CoV-2 to remain viable for hours in particles (van Doremalen et al. 2020), indicates such a transmission pathway is possible. Moreover, considering the high viral loads found in the upper respiratory tract of asymptomatic individuals infected with COVID-19 (Zou et al. 2020), it has been hypothesized that small droplets and particles generated during normal respiratory behaviors, such as breathing and speaking, could lead to the fast spread of the disease (Edwards et al. 2004,

Fabian et al. 2008, and Johnson et al. 2009). However, despite a number of studies of particle generation from these behaviors (Johnson et al. 2009, Papineni and Rosenthal 1997, Chao et al. 2009, Haslbeck et al. 2010, Johnson et al. 2011, and Heo et al. 2017), there is a lack of *in situ* characterization, particularly for breathing due to its low yield of particle production, limiting our ability to model the spread of particles associated with asymptomatic individuals. Specifically, most studies on the size distribution of such particles use devices such as aerodynamic particle sizers (APS) and optical particle counters (OPC) which require transporting the particles to the sensor and do not account for particle evaporation and particle losses during the transport (Bake et al. 2019). The only *in situ* measurement (Chao et al. 2009) utilizes interferometric Mie imaging (IMI), which captures particles above 2 µm with measurement accuracy depending on the assumptions of particle refractive index and shape (assumed to be spherical). It is worth noting that recent *in situ* measurements of speech-generated particles using laser sheet imaging with an iPhone 11 camera (Stadnytskyi et al. 2020) was carried out with high speaking volume (>59 dB) which should not be considered as normal respiratory behaviors. Accordingly, no study has conducted computational fluid dynamics (CFD) simulation of the change of size and concentration of particles over time and their spatial variation in an enclosed environment to provide quantitative assessment of the risk of airborne infection. These models are necessary for producing scientifically driven policy regarding social distancing measures and safe business re-opening.

Therefore, in the current study, we present the first detailed characterization of the particle generation process of normal human breathing by combining quantitative Schlieren imaging and multi-magnification digital inline holography (DIH). Such measurements, conducted with eight participants, provide the instantaneous and ensemble average flow field of exhaled gas as well as the concentration, size, and shape distributions of particles ranging from 0.5-50 µm within it. This information is then used as the inputs for high-fidelity CFD simulation of particle transport under several practical settings, which considers the evaporation, drag, gravity, and residence lifetime of each particle produced by a simulated asymptomatic individual. The simulation results are then used to assess the potential of airborne disease transmission associated with the normal respiratory behaviors under these settings.

## 2. Methodology

In the breathing experiment, a participant is seated and instructed to breath using a nose inhale and mouth exhale at a rate of 15.2 breaths per minute with a 2:3 inhale-exhale ratio, within the range of normal breathing patterns (Tobin et al. 1983). The participate first breathes five times in front of a high-speed Schlieren imaging setup. The acquired Schlieren images are then processed using optical flow method to determine the instantaneous and averaged flow fields of normal breathing (Liu and Shen 2008), from which the instantaneous/averaged volumetric flow rate and spatial extent of the exhaled flow are obtained. Subsequently, digital inline holography (DIH) measurements are conducted to determine the particle generation from breathing. DIH is an optical diagnostic technique which allows in situ imaging of individual microparticles in an extended sample volume (i.e., orders of magnitude larger than conventional microscopy in the imaging depth of field) without focusing (Poon and Liu 2014, Katz and Sheng 2010, and Xu et al. 2014). In our experiment, DIH measurements with both 1x and 20x magnifications are implemented to capture the particles from above and below 5 µm, respectively. The sample volumes of DIH with both magnifications are positioned 1.5 cm away from the mouth to capture in situ the original forms of particles generated directly from breathing with minimal influence of evaporation. For DIH measurement with each magnification, the participant is instructed to breath in the same fashion seating in front of the DIH setup for 20 minutes (30 s breathing alternates with 30 s rest, in total 10-minute breathing data). The DIH data (i.e., holograms) with exhaled particles present in the sample volume are first selected, and subsequently reconstructed using Rayleigh-Sommerfeld diffraction kernel (Katz and Sheng 2014). The reconstructed holograms are then processed using automatic image analysis software validated with manual checking using ImageJ to determine the concentration, size (in terms of equivalent diameter), and shape (in terms of circularity) of particles generated from breathing. The breathing experiment described above is conducted over in total eight participants with varying gender and age. The data from all the participants are compiled to obtain the ensemble-averaged results used as the inputs for the CFD simulations. The simulation is used to model the distribution and accumulation of particles under different practical settings. The CFD simulations are conducted using the OpenFoam-6 platform, with the Eulerian-Lagrangian framework for the gas liquid phase simulation (Jasak et al. 2007). Three practical settings (i.e., elevator, small classroom, and small supermarket) are chosen for the simulation. For each setting, different ventilation conditions are simulated to study the effects of ventilation on the dispersion of the exhalation particles. Considering the room temperature and 40% humidity in this study, the classical evaporation model based on quasi-steady-state assumption is used to account for the evaporation of droplets (Ranz and Marshall 1952). Based on our experimental observation, we also assume all the large droplets will eventually evaporate into residual particles of 1.5 µm, i.e., a threshold corresponding to the peak of our particle size distribution measured in our breathing experiments (representing the most-probable residue size). The detailed reasoning of this assumption is provided in the Results Section and supplementary materials. All the injected particles are tracked by the simulation model described in the supplementary materials. The simulation model stops tracking the particles once they encounter surfaces such as walls and considers the particles deposited on the surfaces. The details of experiment and CFD simulation are provided in the supplementary materials.

## 3. Results

*3.1. In situ measurements of particle generation during normal breathing*

The ensemble average flow field of exhaled gas (Fig. 1A), characterized using quantitative Schlieren imaging, shows an axisymmetric cone shape with an averaged cone angle ($\bar{\theta}$) of 25.0°. The streamwise flow velocity averaged over the cross section of the exhalation cone ($\langle\langle u \rangle\rangle$) decays from 0.3 m/s near the mouth to almost zero at about 200 mm (< 1 feet) downstream. These measurements demonstrate the limited spatial range of direct influence associated with normal breathing, in contrast to violent expiratory behaviors such as coughing which yields a cone angle of 65° (Gupta et al. 2009) and flow speed up to 11.2 m/s with the influence zone extending up to ~2.5 m (Bourouiba et al. 2014). The exhaled gas flow fields from different participants show similar patterns with small variation in quantitative measures (e.g. cone angle, exhaled flow rate, exhaled flow velocity etc., Table S1 with details in the supplementary materials). The normalized exhaled gas flow rate ($\hat{Q}_E$) in an exhalation cycle (period $T_E$) extracted from flow field, rises sharply at the beginning of the cycle, peaks around $0.2T_E$ followed by a sustained rate over a duration of about $0.3T_E$, and then decays rapidly (Fig. 1B). Such breathing pattern is similar across different participants as they are instructed to breath in the same fashion, but the peak value of $\hat{Q}_E$ ($\hat{Q}_{E,max}$) varies substantially among individuals (Fig. S4 in the supplementary materials) potentially due to different natural breathing depths of each individual (Miserocchi and Emili 1976, Ganong 1995, and Benchetrit 2000).

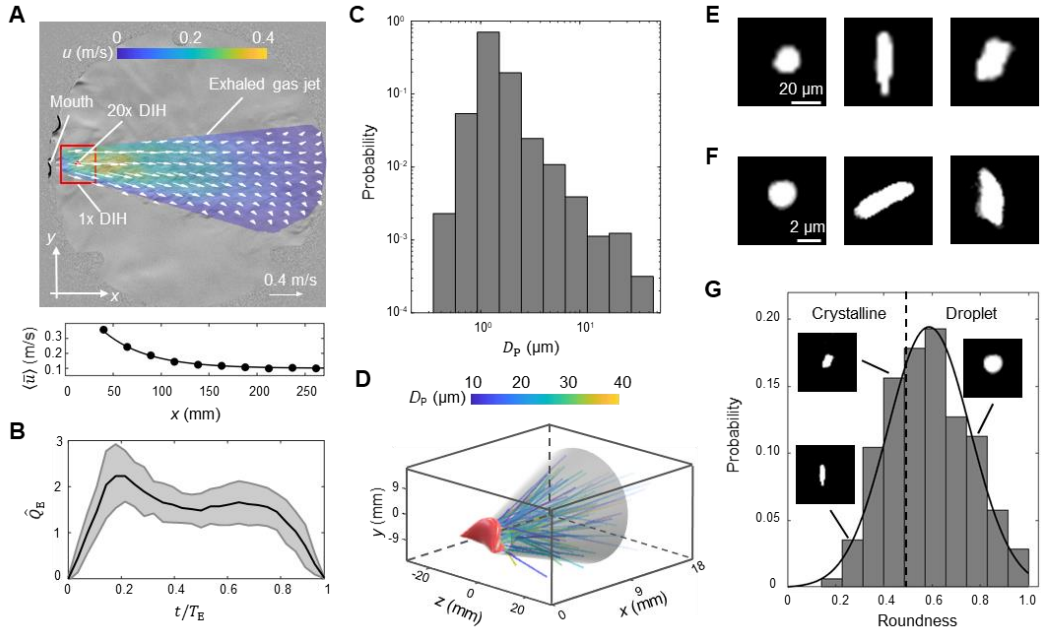

Fig. 1. (**A**) The ensemble average flow field of exhaled gas of all participants superimposed onto an enhanced Schlieren image sample of exhaled gas flow. The details of generating this figure can be found in the supplementary materials. The locations of the mouth, the sampling windows of 1X and 20X digital inline holography (DIH) are marked in the figure. Additionally, the streamwise velocity averaged over the cross section of the exhalation cone ($\overline{\langle u \rangle}$) is plotted against the streamwise distance to the mouth ($x$) to show the decay of flow velocity. (**B**) The change of the normalized exhaled gas flow rate ($\hat{Q}_E$) in one exhalation cycle of time period $T_E$, where $\hat{Q}_E$ is the instantaneous exhaled gas flow rate ($Q_E$) divided by its average for each exhalation cycle. The solid curve and shaded area represent the ensemble average and variance of $\hat{Q}_E$ of all participants, respectively. (**C**) The histogram of particle size quantified using area equivalent diameter ($D_P$). (**D**) The 3D trajectories of all the particles captured using 1X magnification DIH. Sample images of particles from (**E**) 1X and (**F**) 20X DIH measurements. (**G**) The histogram of particle shape quantified using particle roundness (Fig. S7 with details in the supplementary materials) with inset figures showing samples of particles with different roundness levels. The solid line is the fitted normal distribution and the dashed line corresponds to the roundness of 0.5.

The DIH measurements provide the first detailed characterization of the generation of particles during normal breathing in terms of their concentration, size, and shape. The measurements have shown an average concentration of 170 particles per liter exhaled gas (i.e., 44 particles per breath) from an ensemble average of 160 minutes DIH data from eight participants. The particle size distribution peaks around 1.5 µm with a sharp decay towards smaller and larger sizes and has an averaged value of 1.7 µm (Fig. 1C), which is substantially higher than the 0.6 µm obtained from OPC measurement (Papineni et al. 1997) and close to the 2 µm measured using microscopic examination of particles deposited on a glass slide through an impactor (Papineni et al. 1997). Most particles are below 5 µm and only 0.2 % above. The 3D trajectories of particles (>5 µm) within the exhaled gas fit within the breathing cone determined from the Schlieren imaging for all the participants with about 10% particles leaking from the side of the mouth occasionally (Fig. 1D). Besides the particle size, the shape of particles can be obtained using DIH. Interestingly, in addition to a large fraction of round-shaped particles, a substantial fraction yields irregular shapes with edges and corners. Observed both above (Fig. 1E) and below 5 µm (Fig. 1F), these two types of particles with distinct shapes correspond respectively to the droplets and crystalline particles generated from human breathing reported in the literature (Papineni et al. 1997, Morozov and Mikheev 2006). Particularly, through mass spectra of particles, the literature has revealed the presence of nonvolatile solutes such as potassium, calcium, and chorine contents in the crystalline particles (Papineni et al. 1997, and Morozov and Mikheev 2006), which are likely to be generated from the alveolar fluid from the lower respiratory tract (Johnson and Morawska 2009). In addition, Morozov and Mikheev (2006) suggested that these crystalline particles can contain lipids and hydrophobic proteins which are soluble in organic solvents. Noteworthily, such crystalline particles could be the residues of droplet evaporation and may correspond to the long-lasting dry particles reported in the experiments of Stadnytskyi et al. (2020). This information serves as the basis for setting the residue size for droplet evaporation in our simulation (detailed reasoning provided in the supplementary materials). To quantify the content of particles based on their shapes, the histogram of roundness of particles (Fig. S7 with details in the supplemental materials), is obtained (Fig. 1G), and a roundness threshold of 0.5 is selected to categorize the particles into droplet and crystalline types according to the literature (Powers 1953, and Hamilton and Adie 1982). Accordingly, our measurements suggest about 33% of particles produced by normal breathing are crystalline type appearing both below and above 5 µm (Fig. S12 shows that particle roundness is independent with the size). These crystalline particles are usually hygroscopic (Morozov and Mikheev 2006) and could take up moisture from the environment with increasing humidity to form droplets up to 2.5 times of their original sizes (Zieger et al. 2017). It has been suggested moisture can insulate viruses from extreme environments, in favor of their survival during transmission (Tang 2009). Therefore, these crystalline particles may serve as the major virus carriers for airborne transmission since they are likely to the final products of droplet evaporation and can last substantially longer in the air.

*3.2. Variation of particle generation across different participants*

The particle measurements exhibit interesting variability across different participants using the same normal breathing techniques (Fig. 2). Specifically, the concentration of particles larger than 5 µm varies significantly across different individuals (Fig. 2A) while the particles smaller than 5 µm do not show such large variation (Fig. 2B). In particular, the concentration of particles larger than 5 µm produced by one participant (P1) is more than twice the average concentration of the eight participants. Considering particles in this scale range contain higher viral loads (Alonso et al. 2015, and Zhai et al. 2018) and can evaporate rapidly to be airborne, such an individual can be more effective in spreading viruses when asymptomatic. It is worth noting that other studies on particle generation during breathing and speaking also reported the presence of such superemitters, with population percentage ranging from 6% to 25% (Johnson and Morawska 2009, Asadi et al. 2019), consistent with the percentage of such individual (12.5%) observed in our experiments. Superemitters of particles have been related to the superspreaders of infectious diseases in the literature (Edwards et al. 2004, Asadi et al. 2019, Hammer 2020). Remarkably, the percentage of superemitters in our study (though there is large uncertainty due to the small sample size) coincides well with the 10% of superspreading events of COVID-19 observed in preliminary clinical data (Endo et al. 2020), providing further support to the spreading of COVID-19 through particle generation from asymptomatic individuals. In addition, the fraction of crystalline particles varies from 26% to 40% across different participants (Fig. 2C), and shows a strong correlation with the peak of normalized exhaled flow rate ($\hat{Q}_{E,max}$) defined earlier (Fig. 2D). Considering the connection between $\hat{Q}_{E,max}$ and the natural breathing depth of individuals as noted earlier (Miserocchi and Emili 1976, Ganong 1995, and Benchetrit 2000), our results provide strong evidence that the deeper exhalation can lead to the generation of a higher fraction of crystalline particles from the lower respiratory tract as suggested in the literature (Johnson and Morawska 2009).

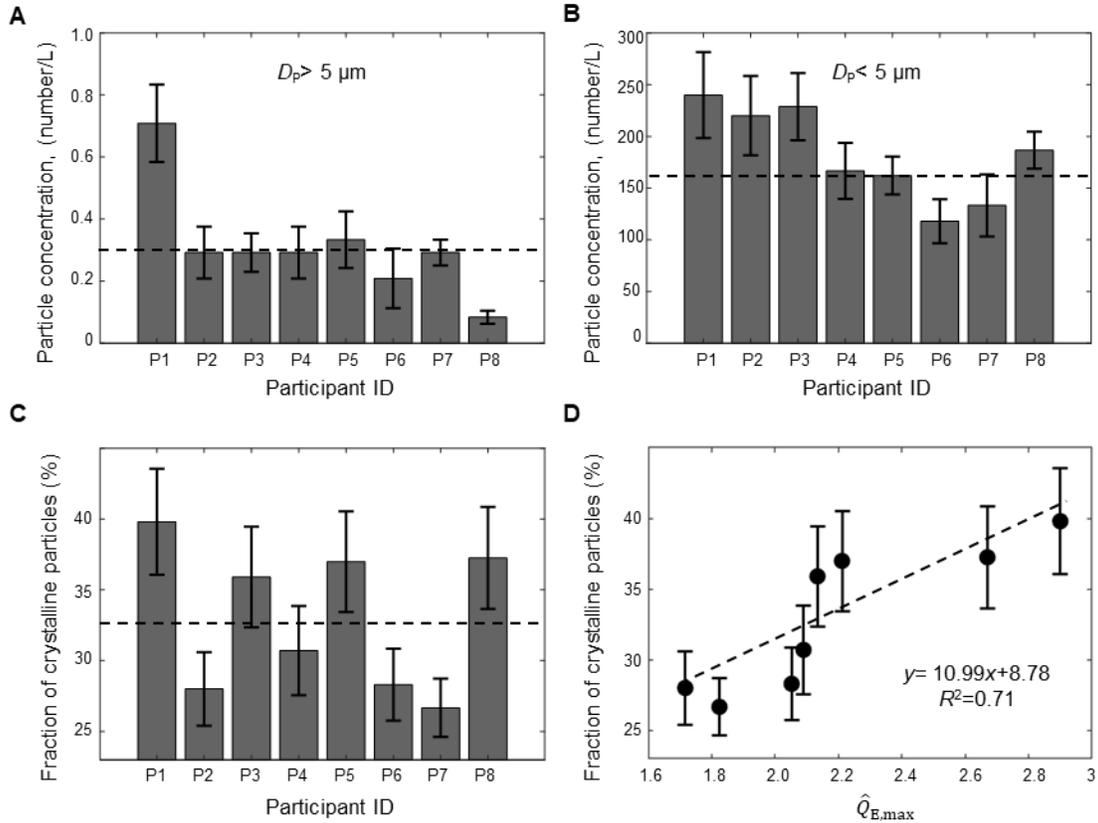

Fig. 2. The variation of particle concentration (number/L) across different participants for (**A**) particles larger than 5 µm and (**B**) smaller than 5 µm. The dashed lines in the figures correspond to the average particle concentration. (**C**) The variation of the fraction of crystalline particles across different participants with its average value marked by the dashed line in the figure. (**D**) The variation of the fraction of crystalline particles with respect to the normalized peak exhaled gas flow rate ($\hat{Q}_{E,max}$) of each participant. The dashed line and the equation in the figure are the least square linear fit of the data.

*3.3. The CFD Simulation of particle transport and deposition under different practical settings*

Using the flow and particle information derived from our breathing experiments, CFD simulations are conducted under three practical settings to determine the particle transport and deposition and assess the risk of being infected ($I_{risk}$) through the airborne transmission of COVID-19 caused by asymptomatic individuals (details of simulation in materials and methods in supplementary materials).

Under the elevator setting (Fig. 3A), a simulated asymptomatic individual (referred to as the "emitter" hereafter) is placed near the wall opposite to the door for one minute. With high ventilation, the particles from the emitter disperse to a large portion of the elevator within one minute, but $I_{risk}$ is extremely low ($\leq 1$) in most of the space (e.g., the two "safe" spots). To assess a riskier scenario, we consider the emitter speaking continuously for one minute and producing particles at a rate 10 times that of normal breathing according to the literature (Asadi et al. 2019).

Without changing ventilation, this scenario exhibits a proportional increase in $I_{risk}$ and the expansion of regions with high risks (e.g., the "hot" spot). With significant reduction in ventilation, the dispersion of particles is confined to one quadrant of the elevator on the emitter side, imposing little risk to the people who are not standing in close proximity to the emitter (e.g., in the two "safe" spots) but two orders of magnitude higher risks for some local hot spots in the quadrant. Remarkably, even under high ventilation, only a small fraction (~15%) of particles is vented out, and the number drops to zero with reduced ventilation. This observation is associated with the presence of stable flow circulation zones in the space (Fig. S13), which traps the particles and increases their residence time to be significantly longer than the simulated time here. In addition, such circulation zone strengthens with increasing ventilation causing more wall deposition of particles (Fig. S16).

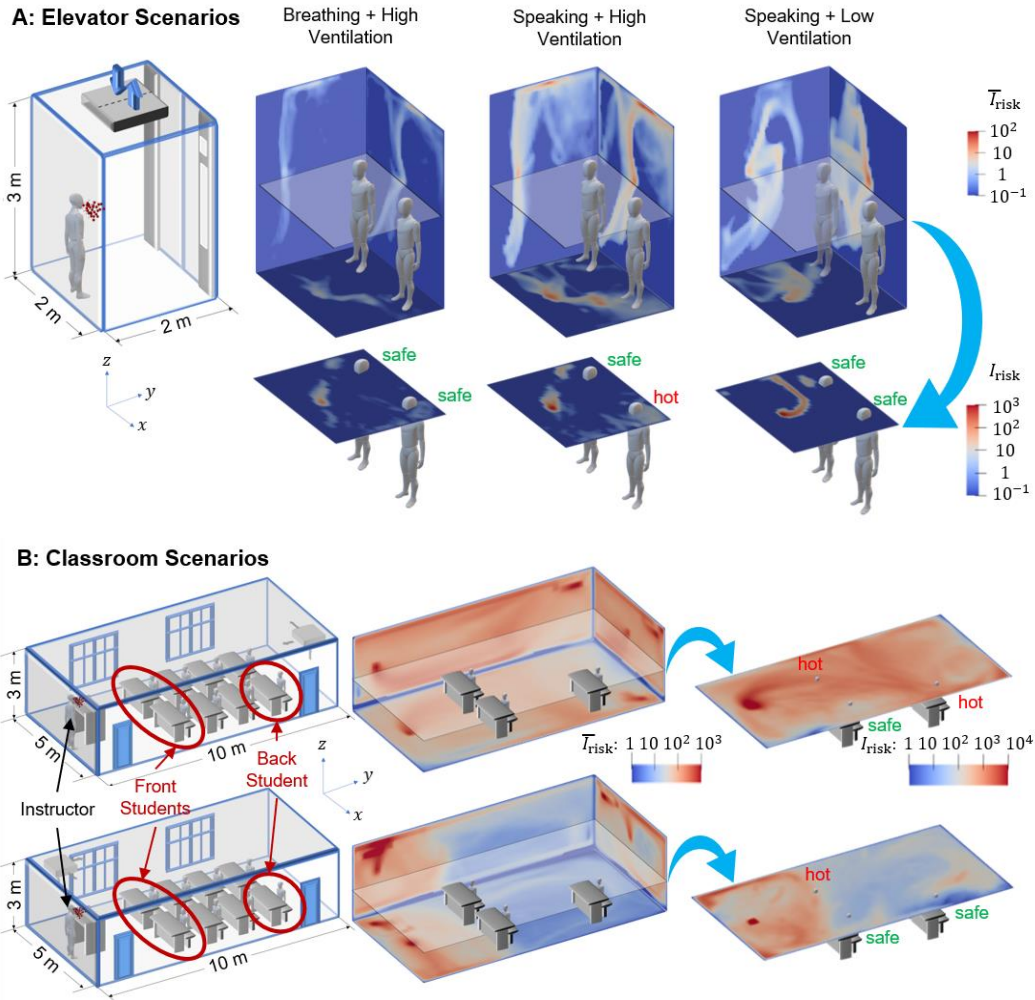

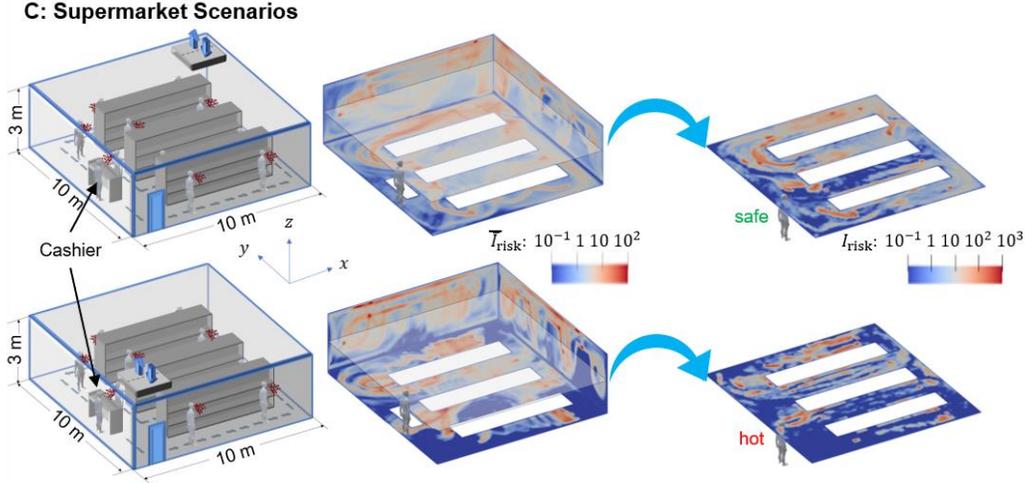

Fig. 3. The CFD simulation of particle transport and deposition to evaluate the risk under (**A**) an elevator setting with a simulated asymptomatic individual breathing under high ventilation of 212 cubic feet per minute (CFM) (supplementary movie S1), speaking under high ventilation (supplementary movie S2) and low ventilation of 15 CFM (supplementary movie S3), (**B**) a small classroom setting with a simulated asymptomatic instructor and the ceiling ventilation system located in the back (supplementary movie S4) and front of the classroom (supplementary movie S5), respectively, and (**C**) a small supermarket setting with a simulated asymptomatic shopper (his/her 10 stops along the dashed route are marked in the schematic) and the ceiling ventilation system located at the back corner (supplementary movie S6) and entrance of the supermarket (supplementary movie S7), respectively. Except for the low ventilation setting in **A**, the ventilation rate of each setting is designed to replace all the air in the space every two minutes, representing the upper bound of recommended ventilation condition of each setting. Under each setting, the risk of a person encountering virus-containing particles at one specific location ($I_{risk}$) is evaluated as the total particle number passing through this location during the simulation time, which can be interpreted as number of particles a person can inhale at this location during the simulated time. Note that such an estimate of the inhaled particle number only provides an estimate of maximum particle encounter since it does not consider the detailed flow processes involved in the inhalation of particles. In the 3D contour plots, the wall contours are the contours of $I_{risk}$ spatially averaged (denoted as $\bar{I}_{risk}$) along $x$, $y$, and $z$ directions, respectively. In addition, assuming the height of each individual is ~1.75 m, selected horizontal slices of $I_{risk}$ contour at the height of human mouth are highlighted to show the $I_{risk}$ for individuals standing (1.6 m for **A** and **C**) or sitting (1.2 m for **B**) at different locations, and several representative locations are marked as safe or hot based on their $I_{risk}$ values (safe criterion: $I_{risk} \leq 1$ for **A**; $I_{risk} \leq 200$ for **B**; $I_{risk} \leq 100$ for **C**).

Under the small classroom setting (Fig. 3B), we consider the emitter to be the instructor upfront and the particles are continuously produced through speaking for 50 minutes (the typical duration of a lecture), representing a much riskier scenario in comparison to one of the breathing students being the emitter. When the ceiling ventilation is at the back corner in the classroom (i.e., far from the emitter), the ventilation spreads particles to the back half of the classroom. Particularly, the region near the vent can yield a significantly higher $I_{risk}$, such that a student sitting in a hot spot in the back could inhale several times more particles than a front student at a safe spot. As the ventilation is relocated to the emitter side, the spread of particles is mostly confined to the region before the front students and the $I_{risk}$ for each student is significantly reduced compared to the former scenario. Remarkably, despite the rate of ventilation set to replace all the air in the space every two minutes for both scenarios, no more than 10% of the total emitted particles are vented out after 50 minutes, although the latter (i.e., ventilation near the emitter) doubles the fraction of vented-out particles of the former (Table S2). Such inefficient particles removal through ventilation is largely associated with the presence of many stable circulation regions in the large space (Fig. S14), which increases particle residence time, causes the majority of deposited to surfaces (i.e., 88% of the total, Table S2), and forms hot spots of surface contamination (e.g., the two ground corners on the same side of the classroom for both ventilation scenarios, Fig. S17).

Under a small supermarket setting (Fig. 3C), the emitter is considered as an asymptomatic shopper with regular breathing, who makes 10 stops (three minutes for each stop with the last stop at the cashier) along a designated zigzag shopping path for 30 minutes. When the ceiling ventilation is in the back corner, the particles spread across the entire supermarket, and particularly, a hot spot is formed in the space between the leftmost shelf and corner near the ventilation due to strong local circulation and entrainment of higher speed channeling flow formed in this space (Fig. S15). However, the cashier, standing near the entrance, is placed in a relatively safe zone. With the ventilation moved to the entrance, the overall spread of particles is reduced, but several other hot spots emerge, including one in front of the cashier increasing his/her risk by about two orders of magnitude. Compared with the classroom setting, the fraction of vented-out particles (~50% for both scenarios, Table S2) is significantly increased here, even at similar ventilation and shorter time duration. Such increase in the particle removal efficiency is primarily attributed to the motion of the emitter which limits the chance of a large fraction of particles becoming trapped in the same stable circulations. Additionally, due to the presence of dividing structures (i.e., shelves), the stable circulation zones reduce in scale (Fig. S15) in comparison to those under the classroom setting, causing less wall deposition except in the back corner near the ventilation (Fig. S18).

## 4. Summary and Discussions

Combining novel *in situ* measurements and CFD simulations, our study provides the first quantitative assessment of risks due to airborne transmission of viruses generated by asymptomatic individuals in a confined space under ventilation. Through integrated quantitative Schlieren imaging and digital inline holography (DIH), our experimental measurements provide a detailed *in situ* characterization of particle generation through normal breathing, including natural breathing flow field, size distribution, concentration, and shapes of particles over a broad range of sizes. The normalized exhaled gas flow rate calculated from breathing flow field measured by Schlieren imaging shows similar pattern across different individuals, despite variation of peak exhaled flow rate, due to different natural breathing depth. Our DIH measurements show that most exhaled particles are below 5 µm with only 0.2% above. The measurements further reveal the presence of two types of particles in the exhaled gas, i.e., droplets and crystalline types, based on particle shapes. The concentration of particles shows large variability across different individuals, with the presence of one potential "superemitter" out of eight participants. Such ratio (12.5%) coincides approximately with the 10% of superspreading events of COVID-19 (Endo et al. 2020), providing further support of transmission of COVID-19 through particles generated by asymptomatic individuals. Our measurements have also indicated that the fraction of crystalline particles of different participants is strongly correlated with the natural breathing depth of the individual. Our simulation results show significant spatial heterogeneity of risks in confined spaces under three practical settings, supporting the interesting observations of COVID-19 infection associated with air conditioning in a restaurant (Lu et al. 2020). Specifically, although ventilation enables the removal of virus-containing particles, it can help spread particles to larger spaces beyond the proximity of asymptomatic individuals. Inappropriate ventilation can also lead to local hot spots with risks that are orders of magnitude higher than other places depending on the relative positioning of particle emitter, ventilation, and space settings. In addition, ventilation can also enhance particle deposition on surfaces causing patched regions with high surface contamination, consistent with the large amount SARS-CoV-2 RNAs extracted from samples collected from hospital floors and air vents (Liu et al. 2020, and Ong et al. 2020). It is also worth noting that these deposited particles (both crystalline particles and droplets) can form highly-resilient microscale residues on surfaces, which may shield the viruses from the influence of environmental changes and contribute to the long-lasting surface infectivity reported in van Doremalen et al. (2020) as suggested in a recent study (Kumar et al. 2020). Particularly, ventilation at a single location, even at the highest rate in the current practice, is highly inefficient at removing particles, due to the presence of relatively stable flow circulation zones in the space and the large amount of particle deposition on surfaces. This result suggests that improvements to air filters alone are not enough to reduce the particle concentration.

Our study can directly lead to practical guidelines and science-driven policy for mitigating the risks of airborne infection of COVID-19 with minimal impact on the economy and social activities, which are critical for the safe re-opening of many businesses. Specifically, our results suggest that optimizing ventilation settings (e.g., adding more sites of ventilation and/or more turbulence to disrupt stable circulation zones) even under the current ventilation capacity can significantly improve the efficiency of particle removal. Adjusting the placement of occupants (e.g., students or cashier in our cases) in the room to avoid hot spots and frequent cleaning of surfaces prone to contamination can reduce the risks. Wearing masks to cut down the source of particle generation can significantly lower the risks of airborne infection. Additionally, our *in situ* characterization of particle generation through breathing shows its large variability and correlation with individual breathing depth, indicating the need for effective risk assessment at an individual level. Our study can be further extended to a broad range of practical settings (e.g., air cabin, restaurant, gym, etc.) with more detailed physics (e.g., exhalation, inhalation flow physics, etc.) and individual characteristics (e.g., exhalation behavior, movement, etc.) as well as more precise HAVC models incorporated to yield more accurate risk assessment under these settings.

**Acknowledgments:** We thank Hongyuan Zhang for helping out with some visualizations of the numerical data. This work is supported by University of Minnesota Rapid Response Grant from Office for Vice President of Research (OVPR); University of Minnesota Institute for Engineering in Medicine (IEM) COVID-19 Rapid Response Grant program, Co-Sponsored by and the Minnesota Robotics Institute (MnRI) and the Clinical and Translational Science Institute (CTSI) through the National Center for Advancing Translational Sciences (NCATS) of the National Institutes of Health (NIH) Award Number UL1TR002494. The content is solely the responsibility of the authors and does not necessarily represent the official views of the National Institutes of Health (NIH).

**Conflict of interest:** The authors declare that they have no known financial interests or personal relationships that could cause a conflict of interest regarding this article.

**Author contributions:** Jiarong Hong devised the entire study. Jiarong Hong, Siyao Shao, Kevin Mallery, and Santosh Kumar designed the experiments. Siyao Shao, Jiaqi Li conducted the experiments. Siyao Shao, Ruichen He, and Jiarong Hong conducted the analysis of all the experimental data. Suo Yang and Jiarong Hong designed the numerical simulations. Dezhi Zhou and Shufan Zou conducted the numerical simulation. Suo Yang, Dezhi Zhou, Shufan Zou and Jiarong Hong conducted the analysis of numerical simulation data. Jiarong Hong, Siyao Shao, and Suo Yang wrote the paper.

**Data and materials availability:** All data is available in the main text or the supplementary materials. All data, code, and materials are hosted at the Data Repository for the University of Minnesota.

# Supplementary Materials

1. Experimental methods

1.1 Participants

Eight healthy participants including five males and three females, and age ranging from 21 to 29 participates this study. The University of Minnesota Institutional Review Board (UMN IRB) approved this study (NO:00009795), and all research was performed in accordance with the relevant guidelines and regulations of the UMN IRB. Written informed consent was obtained from all participants prior to the study participation and all the participants completed a brief questionnaire including their age, gender, and healthy conditions.

1.2 Breathing patterns

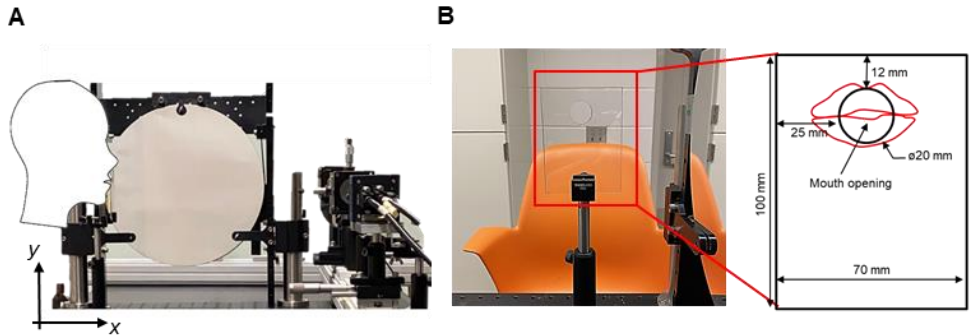

**Fig. S1.** (**A**) Image shown the head position of the participants during the experiments and (**B**) the positioning and dimension of the mouthpiece for aligning the breathing direction of participants.

The rate of participant breathing is set at values that have been identified as normal for healthy individuals (Toobin et al. 1983). The breathing rate is set with the aid of a metronome operating at 76 beats per minute. The nose inhale lasts for 2 beats (1.58 s) while the mouth exhale lasts for 3 beats (2.37 s). A mouthpiece consisting of a 2.0 cm hole in a 6 mm thick acrylic plate is used to align the breathing direction of the participants with the imaging volume. The head position of participants and the dimension of the mouthpiece are shown as Fig. S1. Before the experiments, a presentation describing the detailed breathing patterns was sent to the participants for them to practice the breathing techniques. During the experiments, the participants were instructed to adjust their breathing patterns according to the live view of the Schlieren imaging. This same breathing technique is used for each of the subsequent measurements.

1.3 Schlieren Imaging

Schlieren imaging is a technique that visualizes the index of refraction variation within a fluid that can be caused by variations in temperature, pressure, or composition (Settles et al. 1998). Schlieren imaging has previously been used to characterize exhalation (Xu et at. 2017) and study airflow patterns with and without a facial mask (Tang et al. 2009). The components in the high-speed Schlieren imaging system are shown in Fig. S2. It consists of a blue fiber-coupled LED light source, a concave mirror (30.5 cm diameter, 2.3 m focal length), a blade, and a camera (NAC Memrecam HX-5) with imaging lens (Nikon AF Nikkor 80-200 mm 1:2.8 D). The image size is 960 × 936 pixels (32 × 32 cm field of view) and is recorded at 1000 Hz with a 100 μs exposure time. Each participant takes five breaths for a duration of 20 s. The time-resolved standard deviation of pixel intensity of the Schlieren sequence corresponding to each participant is used to determine the geometry of the exhaled gas jet (i.e., breathing cone angle, shown as Fig. S3). We use optical flow to quantify the flow field, using open source software developed by Liu et al (2008), with a 5-frame skip to ensure sufficient displacement of flow structures inside the exhalation cone. The interrogation window is chosen as 32 × 32 pixels. The flow field is subsequently used to determine the volumetric flow rate of the exhaled gas jet by integration of the flow velocity over the cross section of the exhalation cone at a fixed location (3 cm downstream of the mouthpiece), assuming the jet is axisymmetric. For the ensembled average flow field of the breathing (Fig. 1A in the main text), the averaged flow field of 20 Schlieren video frames around the $0.2T_E$ of each breath was taken first. Then, the flow fields of 40 breaths of all the participants are realigned to a specific breathing cone captured from participant 2 and averaged to have the flow field shown in Fig. 1A. Fig. S4 shows the breathing patterns of all participants observed in the present study. Note that despite similar patterns of breathing, the normalized peak exhaled gas flow rate varies substantially across the participants. Table S1 is a summary of the Schlieren results of all the participants including their average exhaled gas flow rate and the breathing cone angle.

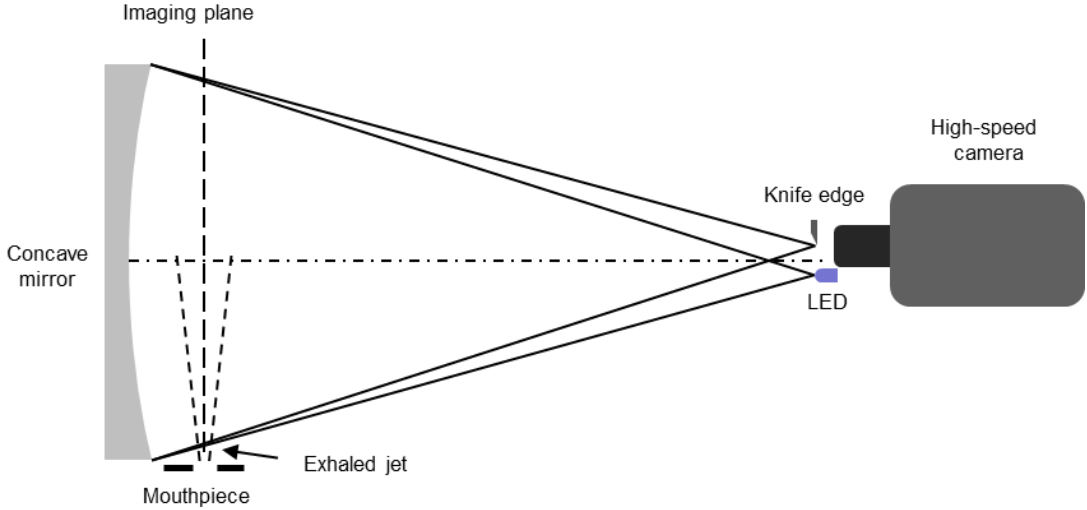

**Fig. S2.** The high-speed Schlieren imaging system.

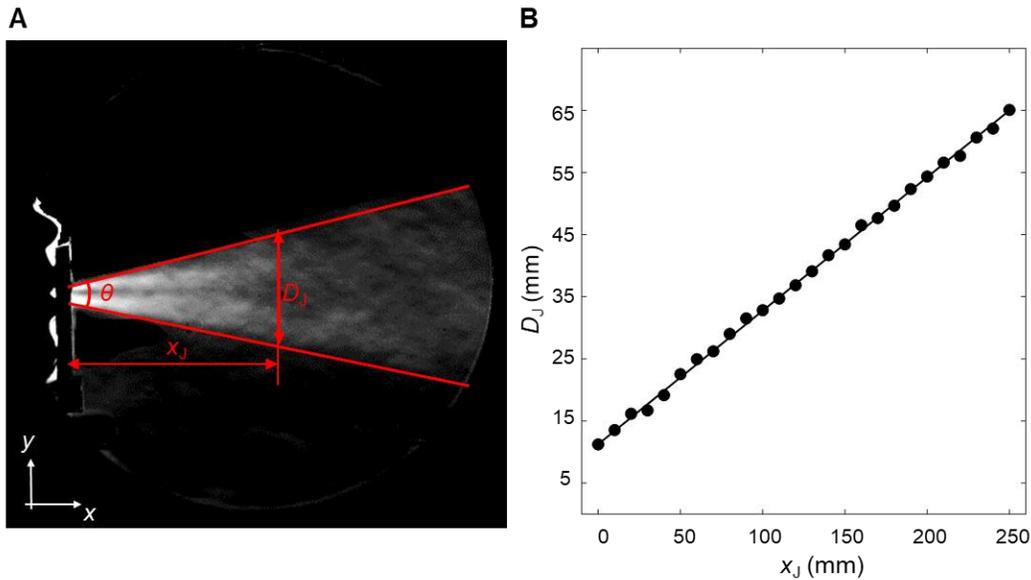

**Fig. S3.** **(A)** Standard deviation of all eight participants to show the extent and geometry of the breathing cone, the breathing cone angle $\theta$ is used to quantify the breathing cone geometry of each participant and **(B)** the linear growth of the breathing cone diameter vs. the distance from the mouthpiece.

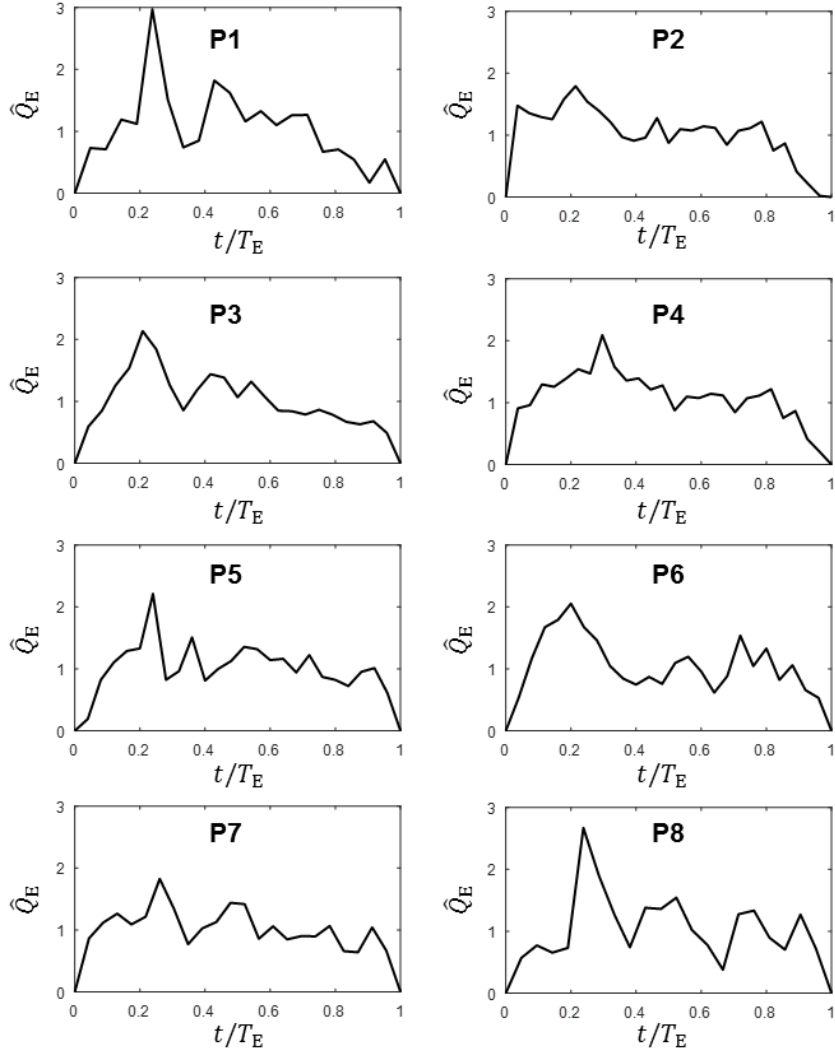

**Fig. S4**. The instantaneous breathing patterns determined by the normalized exhaled flow rate ($\hat{Q}_E$) of all participants. The $\hat{Q}_E$ is the instantaneous exhaled gas flow rate ($Q_E$) divided by its average for each exhalation cycle ($T_E$).

| Participant identification | Schlieren results | | |
|---|---|---|---|
| | Exhaled gas flow rate (L/s) | Averaged flow speed at 30 mm away from the mouthpiece (m/s) | Breathing cone angle (°) |
| P1 | 0.07 | 0.14 | 20.3 |
| P2 | 0.15 | 0.29 | 27.4 |
| P3 | 0.12 | 0.23 | 30.3 |
| P4 | 0.12 | 0.22 | 25.1 |
| P5 | 0.08 | 0.16 | 22.7 |
| P6 | 0.10 | 0.19 | 20.0 |
| P7 | 0.11 | 0.22 | 27.0 |
| P8 | 0.12 | 0.23 | 27.7 |
| Average | 0.11 | 0.21 | 25.0 |
| Standard deviation | 0.02 | 0.04 | 3.8 |

**Table S1.** Summary of the results from high-speed Schlieren measurements of breathing patterns.

### 1.4 1X magnification digital inline holography (DIH) measurements

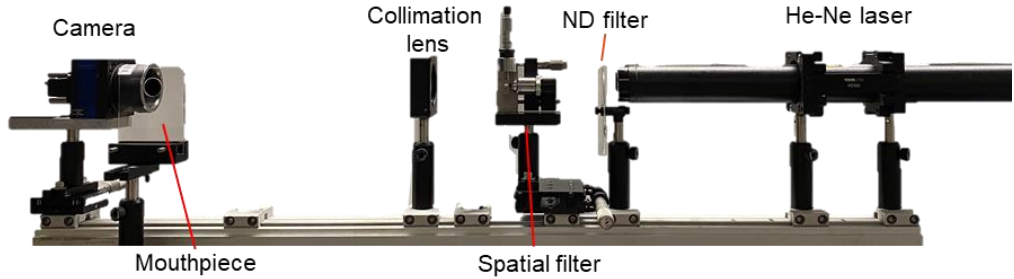

**Fig. S5.** The components of the 1X DIH measurements.

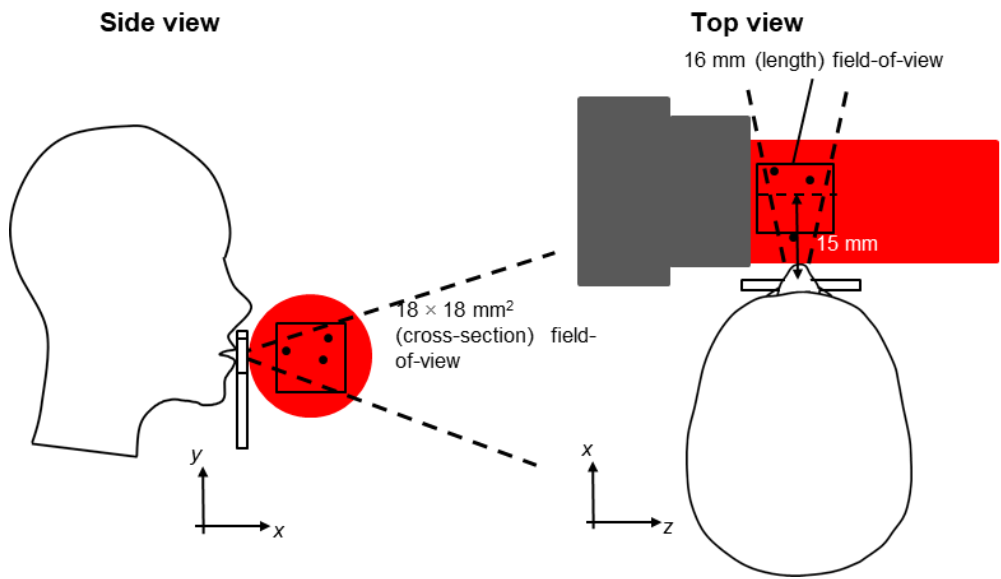

**Fig. S6.** The schematic shown the side view and the top view of the head position and the measurement window of the participant during the 1X magnification DIH measurements of the exhaled particle measurements.

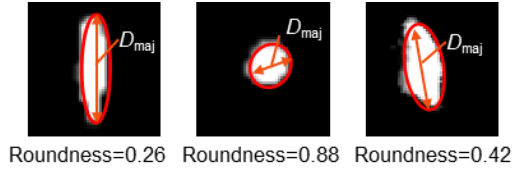

Roundness=0.26  Roundness=0.88  Roundness=0.42

**Fig. S7.** Sample images show the best fitted ellipses of particles their major axis ($D_{\text{maj}}$) used to calculate the roundness of the particles.

DIH is an optical diagnostic technique which allows *in situ* imaging of individual microparticles in an extended sample volume (i.e., orders of magnitude larger than conventional microscopy in the imaging depth of field) without focusing (Poon and Liu 2014, Katz and Sheng 2010, and Xu et al. 2014). DIH operates using the principle of optical diffraction; spatially and temporally coherent reference light beam illuminates the sample (particles). Light scattered by the particles interferes with the unscattered portion of the reference beam. The recorded diffraction pattern (or hologram) can be digitally refocused through convolution with a diffraction kernel, to determine the size, shape, location, and optical properties of the particles. Two magnifications are used to increase the measured size range. The 1X magnification digital inline holographic imaging (DIH) system is used to image exhaled particles ranging in size primarily from 10 µm to 50 µm. Our 1X DIH system, shown in Fig. S5, consists of a He-Ne laser, a spatial filter (Newport Inc) and a collimation lens ($f$=45 mm) producing a 2 cm diameter gaussian beam. A large format CMOS camera (4000 × 4000 pixels; Viewworks Inc) captures the holograms over a 18 × 18 mm field of view at a resolution of 4.5 µm/pixel, sampling at 35 frames/s. Since the exhaled jet diameter is 16 mm, the volume containing particles is 18 × 18 16 mm$^3$ (5 mL). The field of view and center of the laser beam are vertically aligned with the mouthpiece opening and offset by 1.5 cm (see Fig. S6) to ensure the sample volume covers the whole breathing jet at the measurement location. The measurements from each participant are repeated 20 times, with each dataset consisting of a 30 s hologram sequence replicating the breathing technique used in the Schlieren measurements. Hologram processing has three steps. First, the holograms are resized to 1024 × 1024 pixels to reduce processing time and enhanced by time averaged background division to remove stationary artifacts present in the image and remove spatial intensity variations. Then, the holograms are reconstructed by convolution with Rayleigh Sommerfeld diffraction kernel (Katz and Sheng 2010) to obtain a 3D optical field. Such 3D optical field is used to determine the rough 3D centroids of the exhaled particles. The original size (4000 × 4000 pixels) holograms containing particles are reconstructed near the rough longitudinal location determined from the first step to get the precise longitudinal planes of the particles. We combined three criteria to distinguish the exhaled particles from the background particles. First, the particles have a streamwise speed from 0.1 to 0.3 m/s near the DIH measurement window according to Schlieren measurements. Second, the initial $z$ locations (longitudinal) of the particles are restricted between -8 mm and +8 mm to the mouth, which overlaps with the position of the jet from the image plane. Third, the particle trajectories generally follow the exhaled cone shape established from the Schlieren measurements. The particle size and shape are manually measured from 100 × 100 pixels cropped images using ImageJ. We use the area-equivalent diameter $D_{\text{p}}$ (Eqn. 1) to quantify the particle size and use roundness (Eqn. 2) as a measure of particle shape. The cross sectional area and the major axis length of the particles are given by $A$ and $D_{\text{maj}}$, respectively, where the latter is determined by an ellipse fit (a few samples are shown in Fig. S7).

$$D_{\text{p}} = \sqrt{4A\pi} \qquad (1)$$

$$\text{Roundness} = 4A/(\pi D_{\text{maj}}^2) \qquad (2)$$

1.5 20X magnification DIH measurements

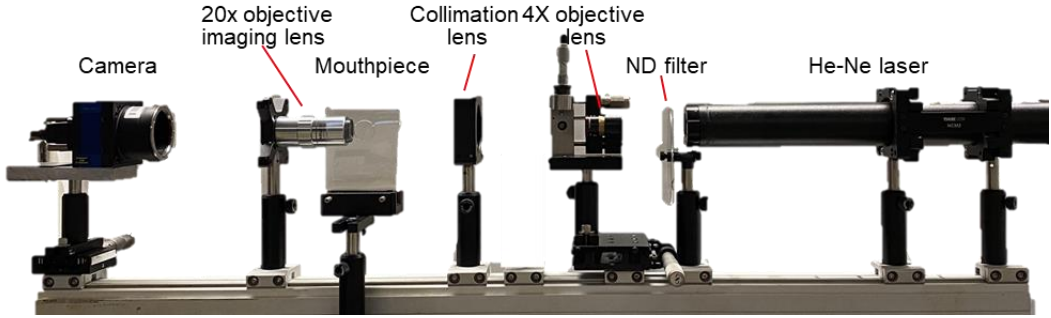

**Fig. S8.** The components of the 20X DIH measurements.

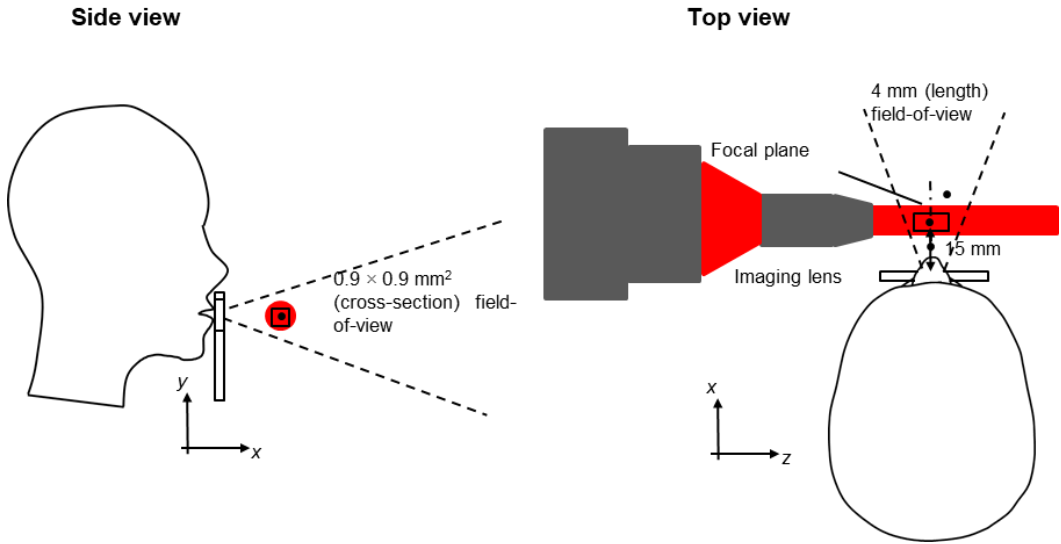

**Fig. S9.** The schematic shown the side view and the top view of the head position and the measurement window of the participant during the 20X magnification DIH measurements of the exhaled particle measurements.

We modify the 1X DIH imaging system by introducing a 20X microscopic objective (Mitutoyo Plan Apo 20X, 0.42 NA) and replacing the spatial filter by a 4X objective lens for reducing the beam expansion to 3 mm from 2 cm. The camera provides an imaging resolution is 0.23 µm/pixel resulting in a sample volume of $0.9 \times 0.9 \times 4$ mm³ ($3 \times 10^{-3}$ mL). The individual components and participant head position for the 20X DIH measurements are shown as Fig. S8 and Fig. S9, respectively. The reduced depth of the sample volume is due to the magnification of the imaging objective lens. The system is capable of resolving particles that are around 0.5 µm very accurately. Similar to the 1X data acquisition, the 20X experiments consist of 20 repetitions with each participant, generating 30 s of data each. Once recorded, the holograms are processed following the same algorithm described above for the 1X DIH measurements.

## 1.6 Combining 1X and 20X DIH measurements

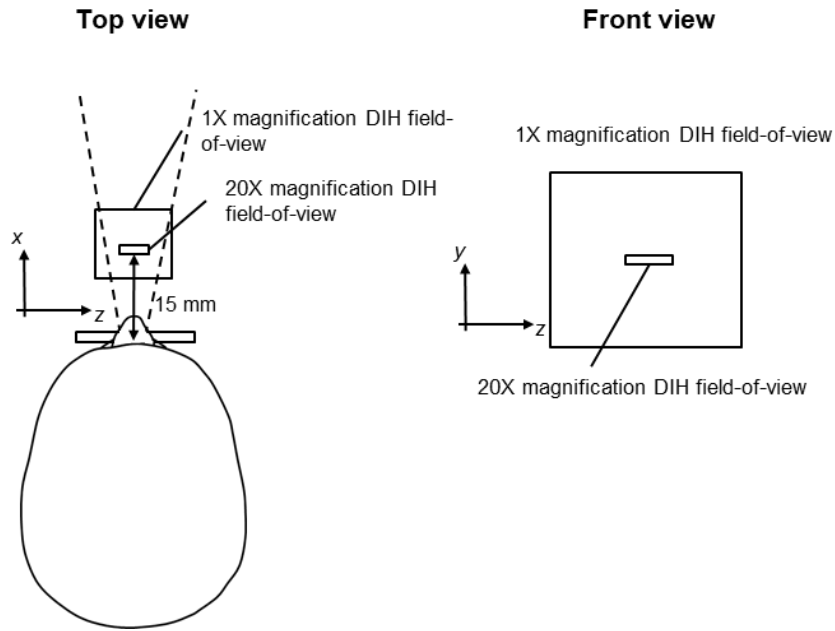

**Fig. S10.** The schematic shown the top view and the front view of the relative positions of 1X and 20X magnification DIH measurements during the experiments.

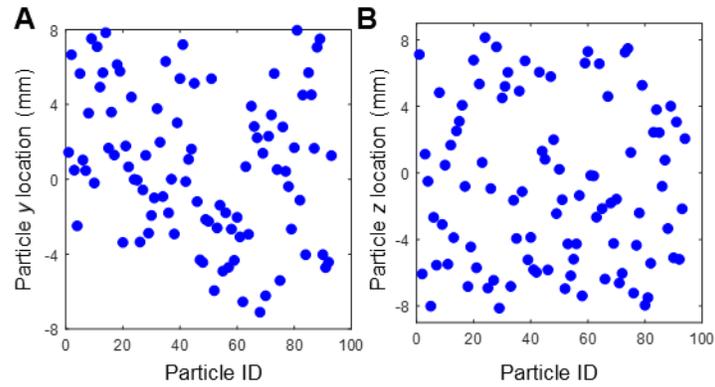

**Fig. S11.** (**A**) The $y$ (vertical) location distribution of the particles captured in 1X measurement and (**B**) the $z$ (horizontal) location distribution of the particles. The origin of the coordinates is set as the center of the mouthpiece.

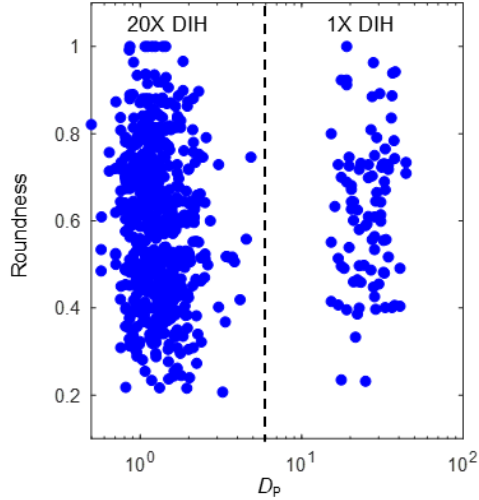

**Fig. S12.** A scatter plot showing the particle roundness versus particle area-equivalent diameter ($D_P$).

To compare the total concentration of the particles measured by both systems, we multiply the volume-averaged count by the ratio of the volumes for both measurements with the assumption that the particles are uniformly distributed in the breathing jet (schematic shown as Fig. S10). We further confirmed such assumptions by plotting out the initial $y$ (vertical) and $z$ (longitudinal) locations of the particles captured from 1X DIH measurements (Fig. S11). Lastly, the particle roundness and particle size are independent with each other as shown in Fig. S12.

2. Numerical Simulation

The simulations are conducted based on the OpenFOAM-6 platform, with the Eulerian-Lagrangian framework for the gas-liquid phase simulation. The gas phase flow is governed by a set of conservation equations:

$$\frac{\partial \rho_g}{\partial t} + \nabla \cdot (\rho_g \mathbf{u}_g) = \dot{S}_m \tag{3}$$

$$\frac{\partial \rho_g \mathbf{u}_g}{\partial t} + \nabla \cdot (\rho_g \mathbf{u}_g \otimes \mathbf{u}_g) = -\nabla p_g + \nabla \cdot \boldsymbol{\tau}_g + \dot{\mathbf{S}}_F \tag{4}$$

where $t$ is time, $\rho_g$ is the density of the gas mixture, subscripts g indicates the gas phase, $\mathbf{u}_g$ is the velocity vector including velocities in three directions. $\boldsymbol{\tau}_g$ is the viscous stress tensor. $\dot{S}_m$ and $\dot{\mathbf{S}}_F$ are the source terms incurred by the Lagrangian particles, which are calculated by:

$$\dot{S}_m = \frac{1}{V_{cell}} \sum_i \dot{m}_{i,d} \tag{5}$$

$$\dot{\mathbf{S}}_F = -\frac{1}{V_{cell}} \sum_i \boldsymbol{F}_{i,d} \tag{6}$$

where the summation over $i$ means the summation over all the Lagrangian particles, $\dot{m}_{i,d}$ is the rate of change in mass of a particular Lagrangian particle, $\boldsymbol{F}_{i,d}$ is the drag force of the Lagrangian particles. The dispersed liquid phase is modeled by a large number of spherical droplets tracked by a Lagrangian model. Due to the small size of the particle in this study, the primary break-up is neglected. In addition, the spray is assumed to be diluted and hence, the interactions between Lagrangian particles are also ignored. The mass rate of change $\dot{m}_{i,d}$ and the drag force $\boldsymbol{F}_{i,d}$ govern the dynamics of each Lagrangian particle by:

$$\frac{\mathrm{d}m_{i,d}}{\mathrm{d}t} = \dot{m}_{i,d} \tag{7}$$

$$\frac{\mathrm{d}u_{i,d}}{\mathrm{d}t} = \frac{\boldsymbol{F}_{i,d}}{m_{i,d}} \tag{8}$$

Only the Stokes drag is considered for the drag force, which is detailed in (Liu et al. 1993). Considering the room temperature and 40% humidity in this study, the classical evaporation model based on quasi-steady-state assumption is used to account for the evaporation because the temperature is reasonably far from the boiling point (Ranz and Marshall 1952). The mass rate of change by evaporation is calculated by:

$$\frac{\mathrm{d}m_d}{\mathrm{d}t} = \rho_d \pi d_d \cdot Sh \cdot D_i \ln\left(\frac{1-X_{i,c}}{1-X_{i,s}}\right) \tag{9}$$

where $d_d$ is the diameter of the droplet, $X_{i,c}$ is the surrounding carrier phase (air in this study) concentration. $X_{i,s}$ is the concentration at the surface of a Lagrangian particle calculated by Raoult's law:

$$X_{i,s} = \frac{X_i p_{\mathrm{sat},i}}{p_c} \tag{10}$$

where $p_c$ is the surrounding pressure approximated by the same method as $X_{i,c}$ and $p_{\mathrm{sat},i}$ is the saturation pressure of the liquid phase (water in this study). $Sh$ is the Sherwood number and $D_i$ is the vapor diffusivity. The Sherwood number and diffusivity are calculated by the Ranz-Marshall model (Ranz and Marshall 1952). All the gas phase differential equations in this section are discretized and numerically solved with the finite volume method using OpenFOAM-6 platform (Jasak et al. 2007).

2.1 Ventilation

In the three different simulation settings (confined spaces including elevator, classroom and supermarket), ventilation is considered as inlet and outlet boundaries in the simulations, which replace the air in the confined space with constant air entrance and exit flow rates.

The ventilator of the elevator case is located at the ceiling, consisting of two adjacent faces with one as the inlet and the other one as the outlet. The vent area of the inlet is 0.5 m². According to the elevator air conditioner ventilation standard retrieved from Quality Elevator Inc. (2001), we consider replacing all the air in the elevator every 2 minutes, which corresponds to a high ventilation flow rate of 212 cubic feet per minute (CFM). As a comparison, the low ventilation case uses 15 CFM as the ventilation flow rate, following the 1970s elevator ventilation standard (National Academy of Science 1997).

For the classroom case, a box-like air conditioner with a size $1 \times 1 \times 0.5$ m³ (the vertical height is 0.5 m) is located at one corner of the domain. Three faces of the air conditioner are attached to the wall, while the other three faces, with the bottom as the outlet and the other two side faces as the inlet, are imposed with fixed value boundary conditions. The ventilation air flow rate is set as 1615 CFM, based on the standard retrieved from Home Ventilating Institute (2010).

In the supermarket case, two scenarios with different ventilation locations are simulated. One location is near the entrance, and another location is in the diagonal corner of the entrance. The ventilation air flow rate is 5297 CFM, corresponding to replacing all the air in the supermarket every two minutes according to Home Ventilating Institute (2010). Ventilation in these simulations is supposed to generate recirculation in the domain, which promotes the airborne transmission of the particles. Considering the ventilation flow rate and the computational domain size, all the cases reach "quasi-steady state" before particle injection after simulation of 60 s.

## 2.2 Particle Injections

The breath frequency, as measured in the experiments, is set as one breath per 4 s. The virus carrier (the liquid parcels) are assumed to be pure water in this study. In the natural breath case, the breath injects particles with the concentration of 44 particles per breath, while with speaking, the number of particles per breath is ten times (440 particles per breath) (Asadi et al. 2019). The person who carries the virus is modeled as a Lagrangian particle emitter, spraying Lagrangian particles into the computational domain. In the elevator cases, the emitter is located at (1 m, 0.5 m, 1.6 m) with the injection direction as (0, 1, 0). The breath injection profile is intermittent: during breath inhale, no particles are injected. We collected data at one minute after the starting of particle injections, representing the typical time duration that a person could stay in an elevator. In the classroom cases, the person emits particles for 50 minutes, with a continuous profile at (2.5 m, 1 m, 1.6 m) with the injection direction as (0, 1, 0), representing the speaking of an instructor for 50 minutes during the period of a lecture. In the supermarket cases, to represent a typical shopping-to-cashier path of a customer, ten stops of the injectors are set, each of which injects particles for a duration of three minutes by normal breathing. We collected the data at 30 minutes, corresponding to that the customer stayed for three minutes at each stop. The last stop is at the cashier.

The injected Lagrangian particles are tracked by the model described in the previous section. We consider a special treatment for the Lagrangian evaporation effect. We assume all the large droplets will eventually evaporate into residual dry particles of finite size. Such assumption is based on the presence of crystalline particles observed in our breathing experiments, which are likely to be the long-lasting dry particles generated by human speaking (Stadnytskyi et al. 2020). Since we cannot conduct *in situ* measurements of the evaporation of respiratory droplets, we set the residue particle size to be uniformly 1.5 µm, i.e., a threshold corresponding to the peak of our particle size distribution measured in our breathing experiments (representing the most-probable residue size). The injected Lagrangian particles are initialized with a size distribution following the experimental results.

## 2.3 Local particle number counting

To illustrate the hot and safe zones in different practical settings and quantify the risk of a person encountering virus-containing particles ($I_{\text{risk}}$) at different locations, we defined an accumulated particle number to represent the number of particles that went through a specific location:

$$P(\pmb{x}) = \sum Pi(\pmb{x}) \qquad (11)$$

where $P$ is the accumulated number of particles, which is a function of the spatial coordinates $\pmb{x}$. To avoid double counting, $Pi$ is defined as:

$$Pi(\pmb{x}) = \begin{cases} 1, \text{ the first time particle } i \text{ appears at location } \pmb{x} \\ 0, \text{ otherwise} \end{cases} \qquad (12)$$

As seen, $P$ represents the total number of the particle passing through a specified location during the whole simulation period. For better interpretation of the spatial distribution of *P*, spatial average values in each dimension are calculated by:

$$P_x(y,z) = \int_{x_{\min}}^{x_{\max}} P(\mathbf{x})\,dx/(x_{\max} - x_{\min}) \tag{13}$$

$$P_y(x,z) = \int_{y_{\min}}^{y_{\max}} P(\mathbf{x})\,dy/(y_{\max} - y_{\min}) \tag{14}$$

$$P_z(x,y) = \int_{z_{\min}}^{z_{\max}} P(\mathbf{x})\,dz/(z_{\max} - z_{\min}) \tag{15}$$

2.4 Discussions on flow streamlines

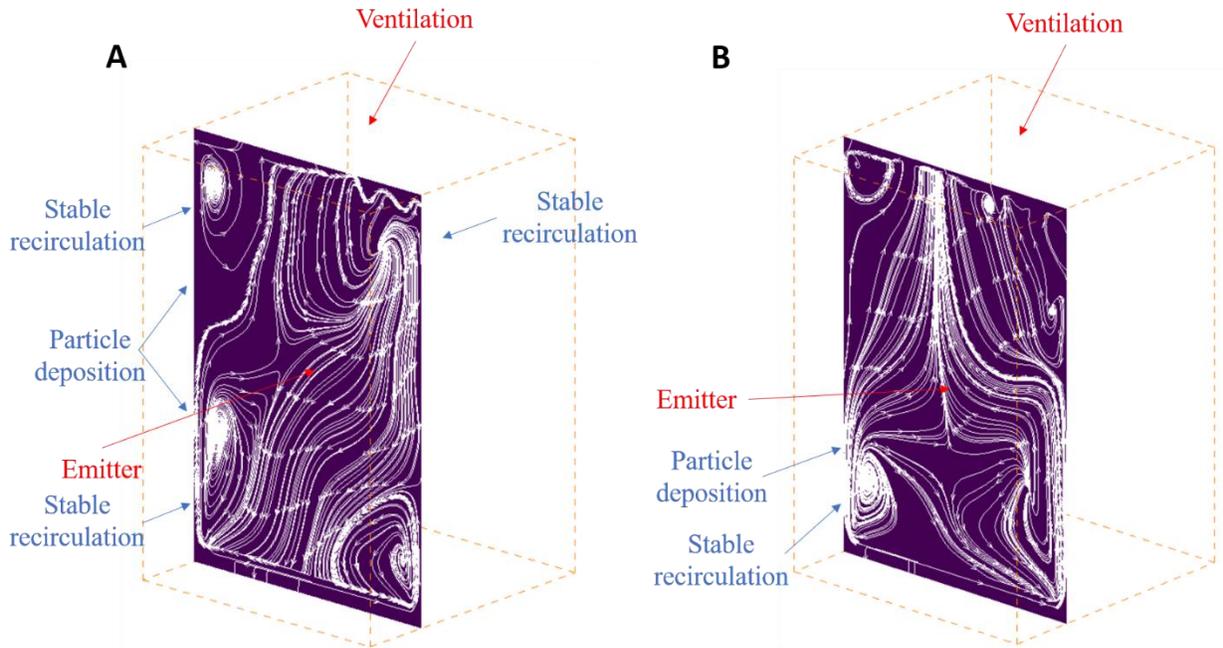

**Fig. S13.** Velocity streamlines on a vertical slice that passes through the emitter for (**A**) ventilation with the high flow rate of 212 CFM and (**B**) ventilation with the low flow rate of 15 CFM. For all cases, many particles are trapped in the stable recirculation zones marked in the plots.

The flow fields with the high ventilation (212 CFM) and low ventilation (15 CFM) flow rates in the elevator are characterized by the velocity streamlines, as shown in Fig. S13. The flow field traps the particles in the recirculation zone, and drives them close to the wall. These particles have a long residence time near the wall, which increases the chance of wall deposition. Weaker recirculation zone is formed with the low ventilation rate of 15 CFM near the roof and thus, most of the particle particles are deposited on the wall by the large recirculation zone near the emitter height, without any particles being removed within the one minute of simulation duration. By contrast, the high ventilation flow rate of 212 CFM is able to suck particle towards the roof within one minute. In the classroom Case B (shown in Fig. S14), the near-instructor ventilation generates a strong recirculation zone near the instructor and keeps the particle from spreading to the far end of the classroom. As a comparison, the far-end ventilation creates a large recirculation throughout the whole classroom, driving the particles to spread towards the far end of the classroom. In the supermarket case (shown in Fig. S15), the valley effect below the far-end (from entrance) ventilation generates a high flow velocity (indicated by the high local density of streamlines) and sucks most of particles to the near-field below the ventilation. However, in the near-entrance ventilation scenario, the particles injected at the first several stops are pulled to the cashier, leading to larger danger of particle encounter (i.e., higher $I_{\text{risk}}$). In addition, as indicated by the density of streamlines, gas flow velocity magnitude when the ventilation is at the far corner is larger than the velocity magnitude when the ventilation is near the entrance, due to the narrow space between the wall and shopping shelf below the ventilation. This explains the enhanced spreading and more homogenous particle distribution when the ventilation is at the far corner.

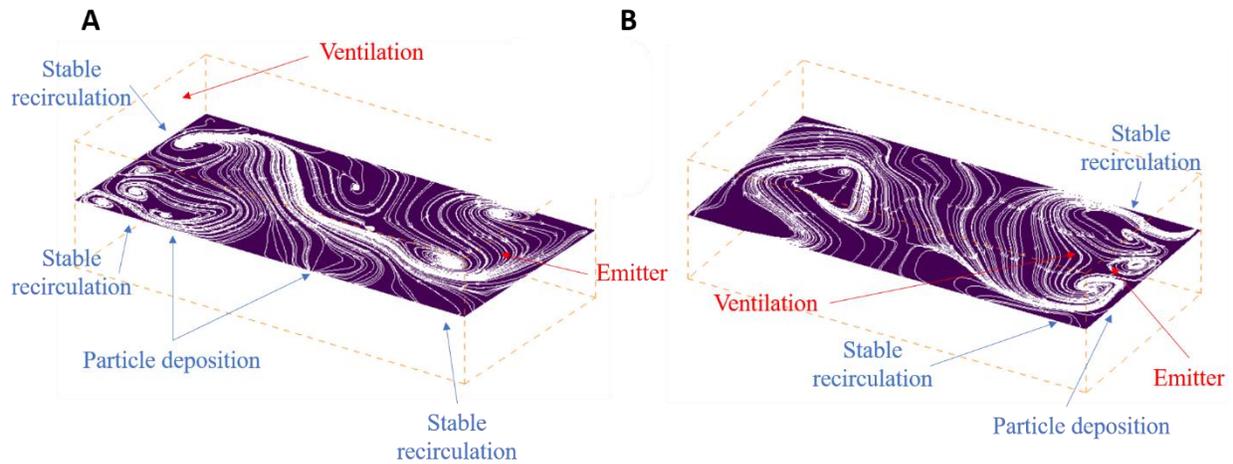

**Fig. S14.** Velocity streamlines at the height of 1.6 m (i.e., the typical height of human noses/mouths) in the classroom cases (**A**) ventilation far away from the emitter (i.e., the instructor) and (**B**) ventilation near the emitter (i.e., the instructor). For both cases, many particles are trapped in the stable recirculation zones marked in the plots.

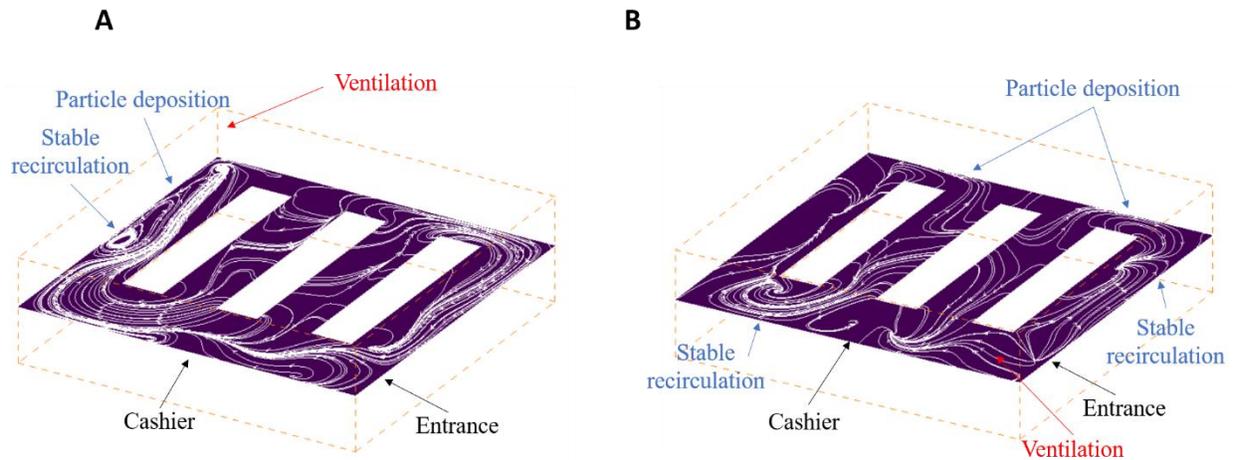

**Fig. S15.** Velocity streamlines at the height of 1.6 m (i.e., the typical height of human noses/mouths) in the supermarket cases (**A**) ventilation far away from the entrance and (**B**) ventilation near the entrance. For both cases, many particles are trapped in the stable recirculation zones marked in the plots.

2.5 Discussions on wall deposition of particles

Compared to the speaking mode, the breathing mode creates less wall deposited particles (Fig. S16) due to the lower number of injected particles. The same ventilation which creates the same flow field, leads to the same wall particle deposition positions on elevator walls. When the ventilation rate is lower, the weak recirculation zone on the top is not capable of spreading the particles to different walls. As a result, the designated flow field drives all the particles to one wall (i.e., the left wall). In the classroom scenario (Fig. S17), the deposited particle numbers are much more than the elevator case, due to continuous speaking of 50 minutes. The comparison between far-end and near-instructor ventilation shows that the near-instructor ventilation keeps the wall particles near the speaking instructor, while in the far-end ventilation case, the created recirculation zone deposits the particles to all the walls in the classroom. In the supermarket settings (Fig. S18), the far-away ventilation sucks a lot of particles to be deposited to the wall near the ventilation, due to the strong velocity magnitude below the ventilation. In the near-entrance ventilation case, particles generated by the far-end shopper directly deposit to the floor due to gravity, since the weak flow field in the far-end field. The distribution of particles (suspended in the air; deposited on the wall; and vented out by the ventilation system) in terms of percentage for each scenario is shown in Table S2.

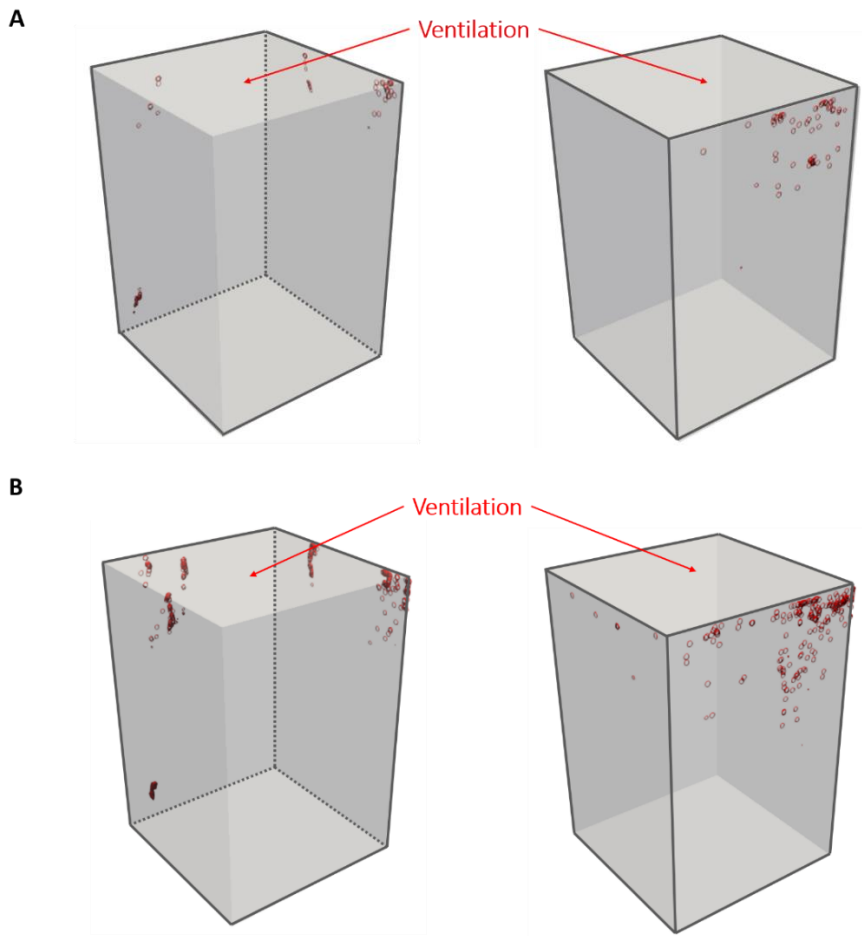

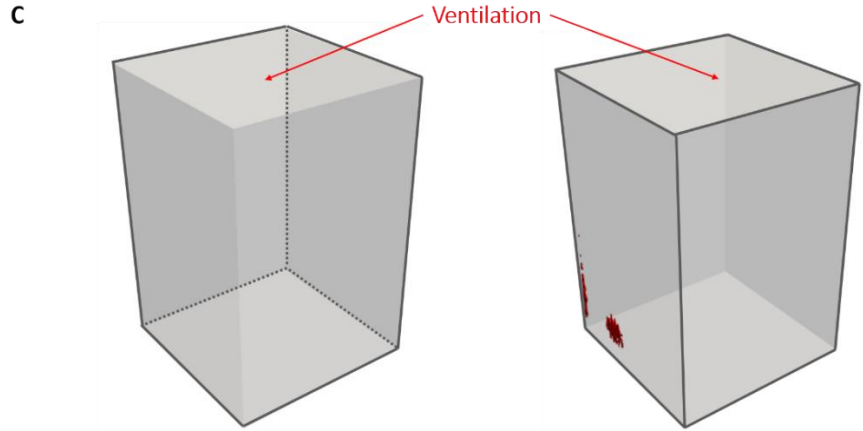

**Fig. S16.** Particle deposition on elevator walls for (**A**) breathing with high ventilation of 212 CFM, (**B**) speaking with high ventilation of 212 CFM, and (**C**) speaking with low ventilation of 15 CFM. For each case, the left subfigure shows the bottom and two inner side walls, and the right subfigure shows the top and two outer side walls.

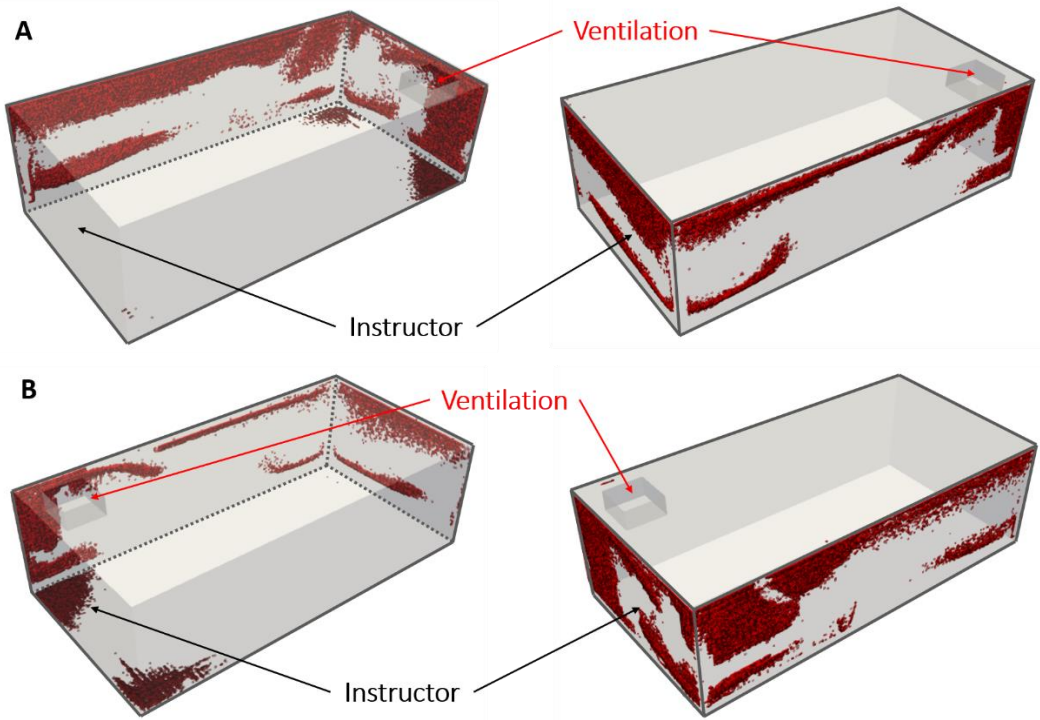

**Fig. S17.** Particle deposition on classroom walls for (**A**) ventilation far from the instructor and (**B**) ventilation near the instructor. For each case, the left subfigure shows the bottom and two inner side walls, and the right subfigure shows the top and two outer side walls.

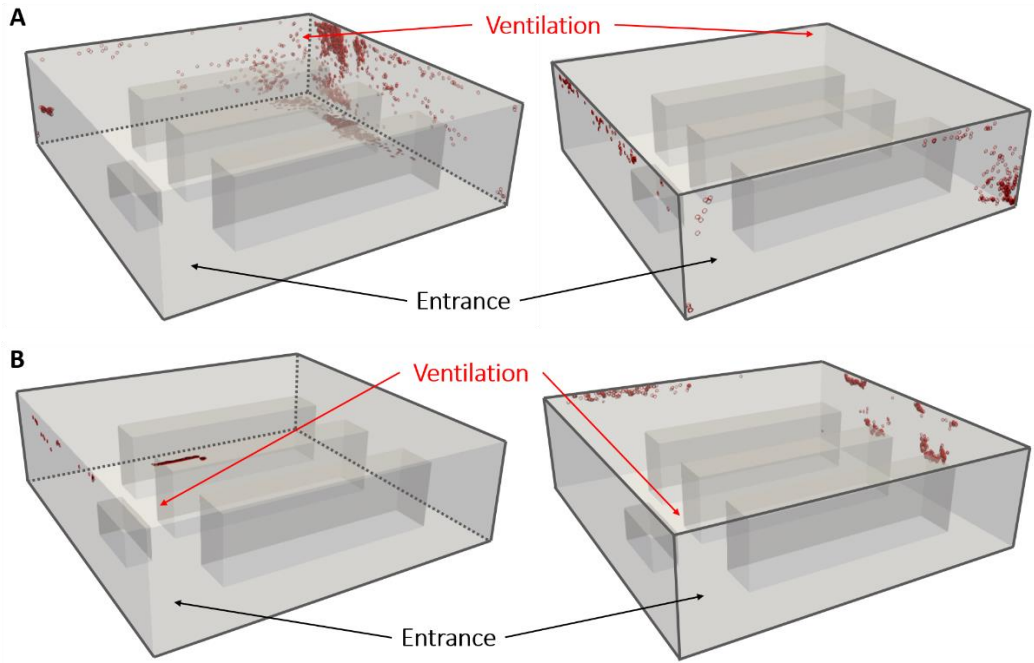

**Fig. S18.** Particle deposition on supermarket walls for (**A**) ventilation far from the entrance and (**B**) ventilation near the entrance. For each case, the left subfigure shows the bottom and two inner side walls, and the right subfigure shows the top and two outer side walls.

| Scenario | | In the air | On the walls | Vented out |
|---|---|---|---|---|
| Elevator | breathing + high ventilation | 78% | 9% | 13% |
| | speaking + high ventilation | 80% | 4% | 16% |
| | speaking + low ventilation | 99% | 1% | 0% |
| Classroom | ventilation far from the instructor | 6% | 88% | 6% |
| | ventilation near the instructor | 2% | 88% | 10% |
| Supermarket | ventilation far from the entrance | 23% | 31% | 46% |
| | ventilation near the entrance | 36% | 12% | 52% |

**Table S2.** Distribution of particles (suspended in the air, deposited on the wall, and vented out by the ventilation system) in terms of percentage for each scenario.